\documentclass[] {article}
\usepackage{pdfpages}
\usepackage{verbatim}
\usepackage[]{graphicx}
\usepackage{amsmath}
\usepackage{amssymb}
\usepackage{epsfig}
\usepackage{color}
 \usepackage{fullpage}

\def\blue{\textcolor{blue}}

\def\##1{{\bf #1}}
\def\=#1{\underline{\underline #1}}

\def\.{\mbox{ \tiny{$^\bullet$} }}

\def\lec{\left\{}
\def\ric{\right\}}

\def\eps{\varepsilon}

\def\lambdao{\lambda_{ 0}}

\def\ko{k_{ 0}}
\def\co{c_{0}}

\def\epssil{\eps_{s}}
  
\def\deg{^\circ}
\def\thetan{\theta_{\rm Can}^{(n)}}
\def\vph{v_{\rm p}}
\def\Deltaprop{\Delta_{\rm prop}}

\begin{document}
\begin{center}
\textbf{Observation of the Uller--Zenneck wave}\\[5pt]

Muhammad Faryad and
Akhlesh Lakhtakia\\[5pt]
{Department of Engineering Science and Mechanics, The Pennsylvania State University, University Park, PA 16802, USA}\\[5pt]
   
\end{center}

{\bf Abstract.} The Uller--Zenneck wave has been theoretically predicted to
exist at the planar interface of two homogeneous dielectric materials of which only one must be
dissipative. Experimental confirmation of this century-old prediction was obtained experimentally
by exciting the Uller--Zenneck wave as a Floquet harmonic of non-zero order at the periodically
corrugated interface of air and crystalline silicon in the 400--to--900-nm spectral regime. Application
for intra-chip optical interconnects at $\sim$850~nm appears promising.\\[10pt]

A doctoral dissertation from 1903 predicted the existence of an  electromagnetic  surface wave (ESW) guided by the planar interface of air and seawater \cite{Uller}, followed four years later by a similar prediction for an ESW guided by the
planar  interface of air and ground \cite{Zenneck}. Both theoretical predictions were made for the radiofrequency (RF) regime, the seawater and the ground being dissipative dielectric materials in that spectral regime. If $\epssil$ is the relative permittivity of 
the dissipative dielectric material such that ${\rm Re}(\epssil)>0$ and
 ${\rm Im}(\epssil)>0$, while $\ko$ is the free-space wavenumber, then the wavenumber $q$ of the Uller--Zenneck wave is given by  
\begin{equation}
\label{dispsol}
q=\ko\sqrt{ { \epssil}/\left({ \epssil+1}\right)}\,,
\end{equation}
where the relative permittivity of air is assumed to be unity. The same formula also holds for surface-plasmon  waves
[${\rm Re}(\epssil)<0$,
 ${\rm Im}(\epssil)>0$]
\cite{SMW} and Fano waves [${\rm Re}(\epssil)<0$,
 ${\rm Im}(\epssil)=0$] \cite{Fano}. All three ESWs are $p$ polarized, but the relative permittivity of the  material partnering air satisfies different conditions for all three.

Controversy has surrounded the Uller--Zenneck surface wave for almost a century \cite{Wise,Wait1998} and RF experiments to excite it on a planar guiding interface have not provided unambiguous proof of its existence \cite{Wait1998,Kukushkin}.  The same ambiguity will prevail in other spectral regimes. However,
if the guiding interface were to be periodically corrugated, theory has recently shown  \cite{FL-UZtheory} that unambiguous proof could be experimentally obtained. As periodically corrugated interfaces are easily fabricated \cite{Lerner} for operation
as diffraction gratings in the optical regime, we decided to experimentally confirm the existence of the Uller--Zenneck wave in this regime using a one-dimensional grating written by electron-beam lithography on a wafer of crystalline silicon. This
communication reports our experimental results.

\begin{figure}[h]
\centering{\includegraphics[width=81mm]{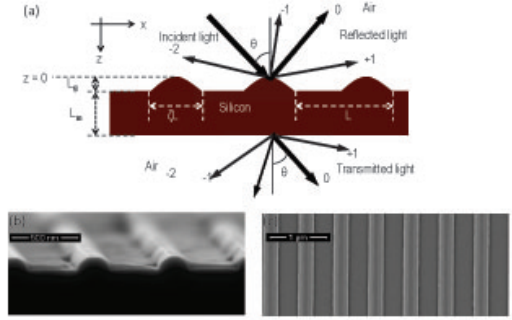}}
\caption{(a) Schematic representation of the experimental configuration used to excite the Uller--Zenneck wave. 
(b)  Cross-sectional and (c) top-view FESEM images of one replicate of the fabricated sample with periodic
corrugations; $L = 600$~nm, $\zeta = 5/12$, and $L_g = 91$~nm.}
\label{Fig1}
\end{figure}

The theoretical foundation of the experiment undertaken is briefly recounted as follows: As shown schematically in Fig.~\ref{Fig1}(a), the regions $z<0$ and $z>L_t=L_g+L_m$ are occupied by air, the region  $L_g< z< L_t$ is occupied by the dissipative dielectric material of relative permittivity $\epssil$, and the region $0< z<L_g$ contains a   grating  of period $L$ along the $x$ axis and  duty cycle  $\zeta \in(0,1)$. Let a $p$-polarized plane wave be obliquely incident upon the grating. The wave vector of the incident plane wave lies wholly in the $xz$ plane and is oriented at an angle $\theta$  with respect to the $z$ axis. The reflected  ($z\leq0$) and the transmitted ($z>L_t$)
fields comprise Floquet harmonics of orders $n\in\mathbb{Z}=\lec0,\pm1,\pm2,...\ric$. Whereas $n=0$ 
 identifies the specular components of the reflected and transmitted fields, the non-specular components are identified by $n\ne0$.  The rigorous coupled-wave approach (RCWA)  \cite{Li93,Moharam95} is useful for computing the reflectances $R_{p}^{(n)}$ and
 the transmittances $T_{p}^{(n)}$ as functions of the angle of incidence $\theta$ and the free-space wavelength $\lambdao=2\pi/\ko$. The absorptance $A_p$ can then be determined using the principle of conservation of 
 energy  \cite{FL-UZtheory}.
 
For   experiments, a $7\times7$ mm$^2$ grating was fabricated  on a 4-inch-thick silicon wafer as follows. The wafer was first spin-coated with ZEP520A photoresist (Zeon, Tokyo) diluted 1:1 with methoxybenzene. Next, the wafer was spun at 2000~rpm for 45~s  and  then baked at 180~$\deg$C for 180~s. The grating pattern was then written using the Vistec 5200 electron-beam lithographic system (Vistec, Best, The Netherlands). Thereafter, the photoresist was developed for 3~min at $-$12~$\deg$C in 
n-amyl acetate, rinsed with isopropyl alcohol (IPA) for 30~s at 20~$\deg$C, and dried using blowing nitrogen. Dry etching was done next on a Versalock 700 system (Plasma-Therm, St. Petersburg, FL,
USA)   for 17~s at   20-mT pressure with chlorine flowing in at  30 sccm. Thereafter, the sample was soaked in PRS-3000 photoresist stripper (Mallinckrodt Baker, Phillipsburg, NJ, USA)
at 85~$\deg$C for 30~min and ultrasonicated  for 120~s. Finally, the sample was rinsed first in IPA for 30~s  and then in de-ionized water for 2~min, before being blow-dried with nitrogen. Two replicates of the sample were made simultaneously.
Images of both replicates were collected on a Leo 1530
field-emission scanning electron microscope (FESEM) (Carl Zeiss, Oberkochen, Germany). Cross-sectional and  top-view FESEM images of one replicate shown in Figs.~\ref{Fig1}(b) and (c) indicate that
$L = 600$~nm, $\zeta = 5/12$, and $L_g = 91$~nm. 

A custom-made variable-angle spectroscopic system was used to measure the specular reflectance $R_{p}^{(0)}$ for
$\theta\in[7\deg,60\deg]$ and $\lambdao\in[400,900]$~nm. In this system, the sample is mounted on a rotatable stage,
and light from a  HL-2000 tungsten halogen lamp
(Ocean Optics, Dunedin, FL, USA) is
passed through a GT10 polarizer (Thorlabs, Newton, NJ)  before hitting the grating such that the incident magnetic field would be parallel to the grating lines. The specularly reflected light
travels through a DH1M   wire grid polarizer (ThorLabs) and is collected using a HRS-BD1-025 CCD spectrometer (Mightex, Pleasanton, CA, USA). The collection time was set at $10$~ms. The following intensities of light were measured:
(i) $I_{\rm dark}$ with the light source switched off and the sample absent;
(ii) $I_{\rm ref}$ with the light source switched on and the sample absent;
and
(iii)  $I_{\rm p}$ with the light source switched on and after specular reflection from the grating.
The specular reflectance $R_{p}^{(0)}$ was then computed as
\begin{equation}
R_{p}^{(0)}=\left({I_{\rm p}-I_{\rm dark}}\right)/\left({I_{\rm ref}-I_{\rm dark}}\right)\,.
\end{equation}
The silicon wafer was extremely thick and so dissipative that $T_{p}^{(n)}=0\,\forall {n}\in\mathbb{Z}$ for $\lambdao\in[400,900]$~nm, which was verified experimentally as well.  

\begin{figure}[h]
\centering{\includegraphics[width=40mm]{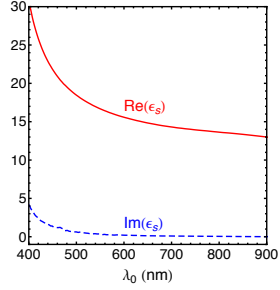}}
\caption{The real and imaginary parts of the relative permittivity $\epssil$ of crystalline silicon
\cite{PalikData} used for computations. 
}
\label{Fig2}
\end{figure}

An ESW can be excited as a Floquet harmonic of order $n$ by the $p$-polarized incident light when $\theta=\thetan$, where
\begin{equation}
\sin\thetan={\rm Re}\lec \sqrt{ { \epssil}/\left({ \epssil+1}\right)}\ric-n{\lambdao}/{L}\,.
\end{equation}
Theory  indicates that
the consequent signature of the excitation of the Uller--Zenneck wave
is a sharp dip at $\thetan$ in the plot of $R_{p}^{(0)}$ versus $\theta$ for constant $\lambdao$ \cite{FL-UZtheory}.
The angles $\thetan$ were calculated as functions of $\lambdao$, with the wavelength-dependent
values of $\epssil$ provided in Fig.~\ref{Fig2}.

\begin{figure}[h]
\centering{\includegraphics[width=81mm]{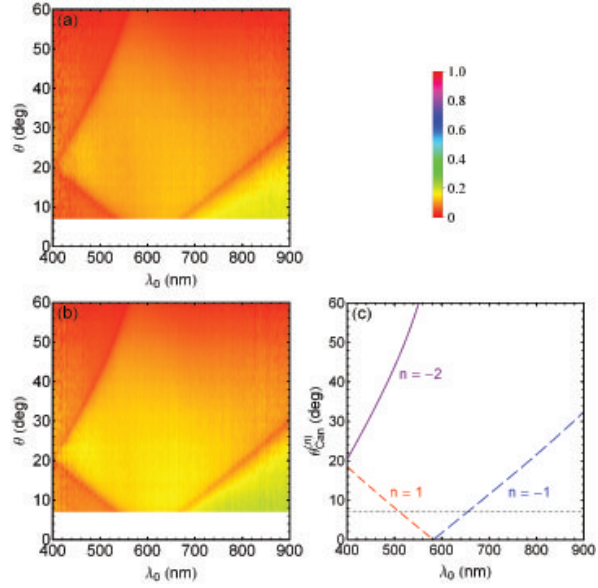}}
\caption{$R_{p}^{(0)}$ of (a) the first and (b) the second replicates measured as  functions of $\theta$ and $\lambdao$.
(c) $\thetan$ for $n\in\lec\pm1,2\ric$ as  functions of  $\lambdao$.}
\label{Fig3}
\end{figure}

Figures~\ref{Fig3}(a) and (b) show $R_{p}^{(0)}$ of the two replicates measured as  functions of $\theta$ and $\lambdao$, while
$\thetan$ for $n\in\lec\pm1,2\ric$ are plotted in Fig.~\ref{Fig3}(c) as  functions of  $\lambdao$. A comparison
of these figures clearly shows that the Uller--Zenneck wave is excited as a Floquet harmonic of
\begin{itemize}
\item[(a)] order $1$ for $7^\circ\leq \theta\lesssim20^\circ$ and $\lambdao\in[400,560]$~nm,
 \item[(b)] order $-2$ for $20^\circ\lesssim \theta\lesssim45^\circ$ and $\lambdao\in[400,570]$~nm,
 and
\item[(c)] order $-1$ for $7^\circ\leq \theta\lesssim31^\circ$ and $\lambdao\in[600,900]$~nm
\end{itemize}

\begin{figure}[h]
\centering{\includegraphics[width=81mm]{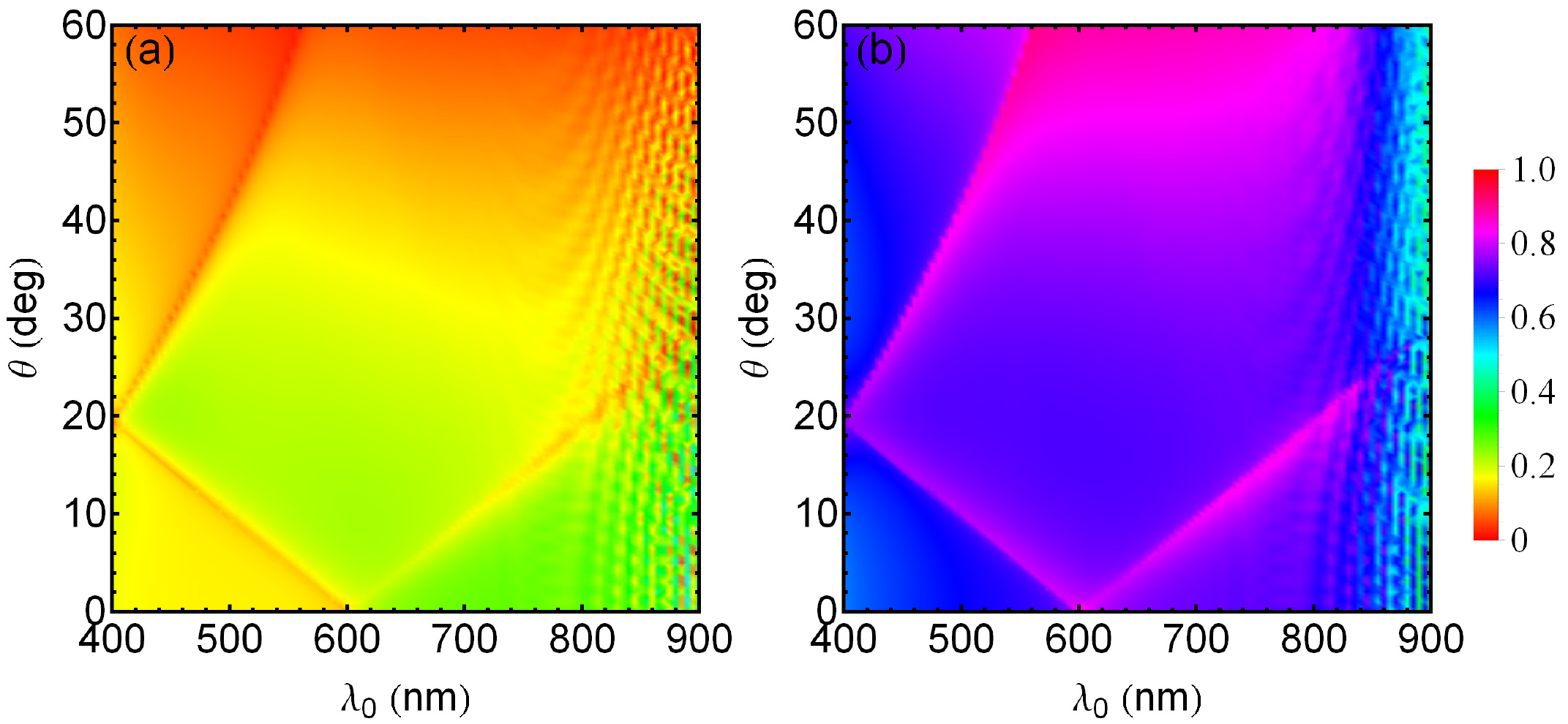}}
\caption{(a) $R_{p}^{(0)}$ and (b) $A_p$ calculated as  functions of $\theta$ and $\lambdao$ for
$L = 600$~nm, $\zeta = 5/12$,  $L_g = 91$~nm, and $L_m=27$~$\mu$m.}
\label{Fig4}
\end{figure}

Further confirmation was provided by RCWA calculations made with
$L = 600$~nm, $\zeta = 5/12$,  $L_g = 91$~nm, and $L_m=27$~$\mu$m. The silicon bump in every period
of the grating was approximated as 
a part of a sinusoid.
Figure~\ref{Fig4} shows the calculated values of $R_p^{(0)}$ and $A_p$ as functions
of $\theta$ and $\lambdao$. The sharp dips in the experimental plots of the specular
reflectance [Figs.~\ref{Fig2}(a) and (b)]
are mirrored as the sharp dips in the analogous theoretical plot  [Fig.~\ref{Fig4}(a)] as well as
the sharp peaks in the theoretical plot of the absorptance [Fig.~\ref{Fig4}(b)]. Parenthetically,
the striations on the right sides of Figs.~\ref{Fig4}(a) and (b) occur due to Fabry--Perot resonances
because ${\rm Im}(\epssil)\simeq0$ for $\lambdao\in[800,900]$~nm.

\begin{figure}[h]
\centering{\includegraphics[width=81mm]{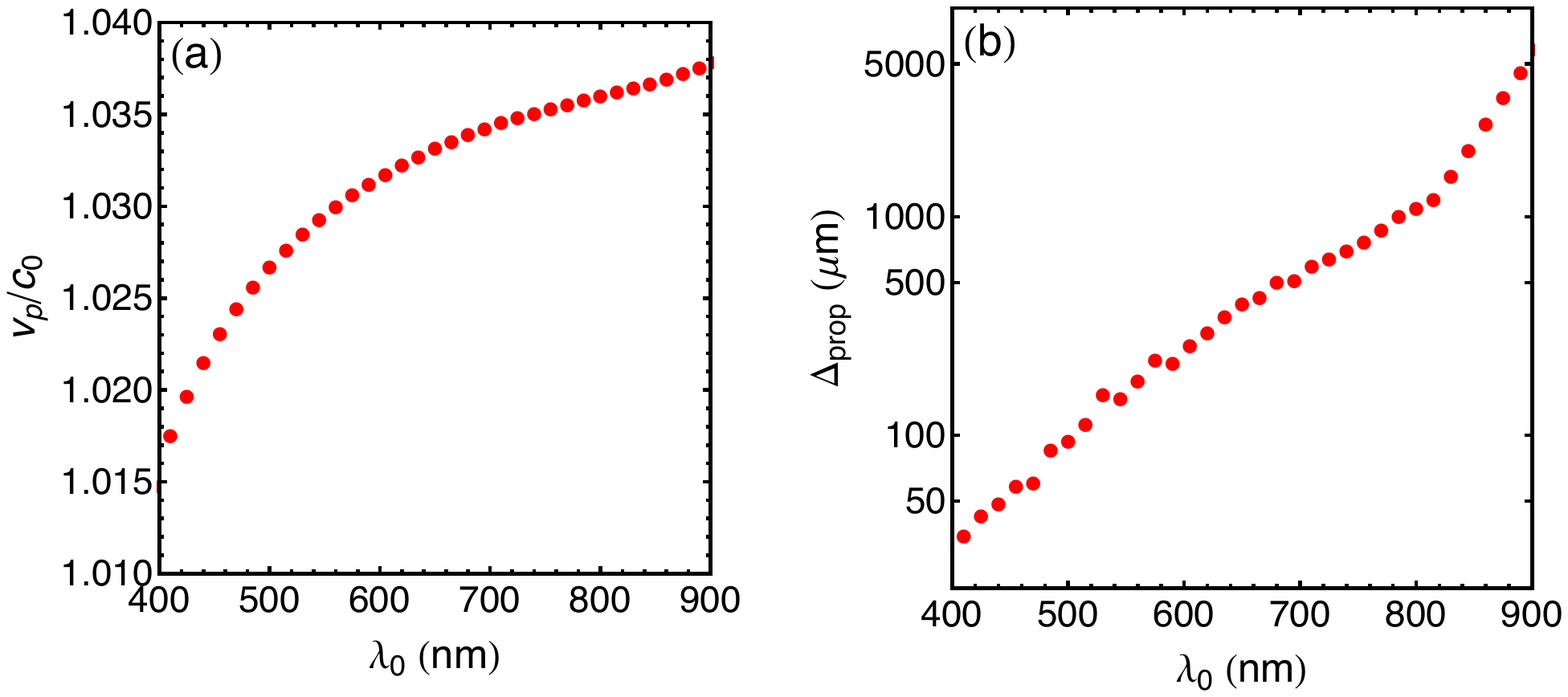}}
\caption{(a) $\vph$ and (b) $\Deltaprop$ calculated as  functions of  $\lambdao$.}
\label{Fig5}
\end{figure}

Having thus experimentally confirmed the existence of the Uller--Zenneck wave, let us also speculate on a
potential use of this phenomenon in the optical regime. For the interface of air and crystalline silicon,
the phase speed $\vph =\ko\co/{\rm Re}(q)$ and the propagation length $\Deltaprop=1/{\rm Im}(q)$
are plotted in Fig.~\ref{Fig5} as functions of  $\lambdao$, where $\co$ is the speed
of light in free space. Clearly, $\vph >\co$, with the excess of $\vph$ over $\co$ increasing  to  $\sim$3.7\% 
at $\lambdao= 850$~nm. At the same wavelength, $\Deltaprop=2.2$~mm, which is a significant distance
in the context of silicon chips for microelectronics. Vertical-cavity surface-emitting lasers (VCSELs)
commonly operate  at $\lambdao\sim 850$~nm, and can be modulated with frequencies in the GHz range \cite{Iga2008,Kaneko}.
Thus, $\sim$850-nm intra-chip optical interconnects could be enabled by the Uller--Zenneck wave.
Regrettably, a similar strategy will not work for silicon photonics which operates in a spectral regime
($\lambdao\gtrsim1500$~nm) in which silicon has minuscule dissipation. But dissipation (i.e.,
${\rm Im}(\epssil)>0$) is essential to the existence of the Uller--Zenneck wave.

\noindent{\it Acknowledgments.} 
This work was supported by the U.S. National Science Foundation (Grant No. DMR-1125590) and the Charles Godfrey Binder Endowment at Penn State.  Thanks are  due to Michael Labella III, Chad M. Eichfeld, and Julie Anderson of the Penn State Materials Research Institute  for assistance with dry-etching, e-beam lithography, and scanning electron microscopy, respectively.

\end{document}